\documentclass[twocolumn,floatfix,showpacs,prb]{revtex4}

\usepackage[]{amsmath}
\usepackage[]{amssymb}
\usepackage[]{amsfonts}
\usepackage[]{graphicx}
\usepackage{rotating}

\newcommand{\ket}[1]{\left|#1\right>}
\newcommand{\kb}[2]{\left|#1\right>\left<#2\right|}

\newcommand{\bok}[3]{\left<#1\right|#2\left|#3\right>}

\begin{document}

\title{Deterministic quantum state transfer from an electronic charge qubit to a photonic polarization qubit}
\author{L. J. P. Ament}
\affiliation{Instituut-Lorentz, Universiteit Leiden, P.O. Box 9506, 2300 RA Leiden, The Netherlands}
\author{C. W. J. Beenakker}
\affiliation{Instituut-Lorentz, Universiteit Leiden, P.O. Box 9506, 2300 RA Leiden, The Netherlands}
\date{February 2006}

\begin{abstract}
Building on an earlier proposal for the production of polarization-entangled microwaves by means of intraband transitions in a pair of quantum dots, we show how this device can be used to transfer an unknown single-qubit state from electronic charge to photonic polarization degrees of freedom. No postselection is required, meaning that the quantum state transfer happens deterministically. Decoherence of the charge qubit causes a non-monotonic decay of the fidelity of the transferred state with increasing decoherence rate.
\end{abstract}


\maketitle

Quantum state transfer between matter and light is a key step in the development of scalable quantum networks. A qubit $\alpha \ket{0} + \beta \ket{1}$ is encoded in matter degrees of freedom for the purpose of computation, and then transferred to photonic degrees of freedom for transportation to a distant location (where it might be converted back to matter for storage or further processing). A promising scheme\cite{Cir97} to accomplish this in the context of atomic physics uses a laser beam to transfer the internal state of an atom to the optical state of a cavity mode. The matter qubit in this case is a superposition of two degenerate ground states of the atom and the photonic qubit is the superposition of an occupied and an empty cavity mode. An all-electronic analogue of this scheme, to transfer a state from one matter qubit to another, has been proposed as well.\cite{Sta04}

In the context of semiconductor quantum dots there exist several proposals for the transfer of a quantum state from electron spin degrees of freedom ({\em spin qubit}) to photon polarization degrees of freedom.\cite{Cer05,Tit05} A separate line of investigation in this context involves a {\em charge qubit},\cite{Hay03,Gor05} i.e.\ a single-electron state $\alpha \ket{A} + \beta \ket{B}$ delocalized over a pair of quantum dots $A$ and $B$. The coupling of a charge qubit to a photon cavity mode was investigated in Ref.\ \onlinecite{Ema05}, as a way to produce polarization-entangled photon pairs at microwave frequencies. Here we build on that proposal to show that the combination of a microwave resonator and three quantum dots can be used to transfer an arbitrary single-qubit state from electron charge to photon polarization degrees of freedom.

\begin{figure}[htp]
	\centerline{\includegraphics[width=8cm]{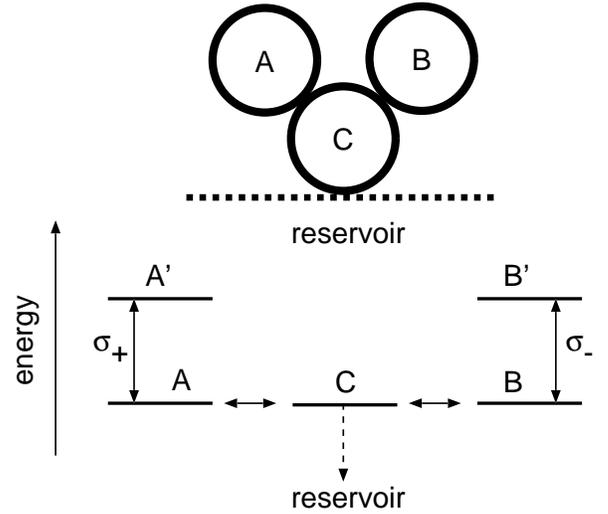}}
	\caption{ Schematic of the model. The upper panel shows a top view of three quantum dots $A,B,C$ connected to an electron reservoir, the lower panel shows the resonant energy levels in the quantum dots. An electron can tunnel between dots $A$ and $C$ or between dots $B$ and $C$ (solid arrows), but not directly between dots $A$ and $B$. From dot $C$, the electron can tunnel into the reservoir (dashed arrow), while the reverse process is prevented by a large bias voltage. A radiative transition within dots $A$ or $B$ is accompanied by the emission or absorption of a photon, with respectively left ($\sigma_{+}$) or right ($\sigma_{-}$) circular polarization. The coupled electron-photon dynamics transfers the charge qubit $\alpha \ket{A'}+\beta \ket{B'}$ to the photon qubit $\alpha \ket{+}+\beta \ket{-}$.}
	\label{fig_fivelevels}
\end{figure}

The device for quantum state transfer, shown schematically in Fig.\ \ref{fig_fivelevels}, differs from the photon entangler\cite{Ema05} only in that it produces a single photon rather than a photon pair. The resonant transitions involve a total of five electron levels in three quantum dots: three ground states $A,B,C$ and two excited states $A',B'$. The radiative transitions $A'\leftrightarrow A$ and $B'\leftrightarrow B$ are resonant with a cavity mode. As explained in detail in Ref.\ \onlinecite{Ema05}, the confining potential and magnetic field can be arranged such that the transition in dot $A$ couples only to left circular polarization $\sigma_{+}$ and the transition in dot $B$ couples only to right circular polarization $\sigma_{-}$.

The charge qubit is prepared initially in the state
\begin{equation}
	\ket{\Psi_{\rm in}} = \left( \alpha \ket{A'} + \beta \ket{B'} \right) \ket{0},\label{Psiindef}
\end{equation}
where $\ket{0}$ denotes the photon vacuum. This single-electron state in dots $A$ and $B$ can decay into a reservoir via a third dot $C$, leaving behind a photon in the cavity. The quantum state transfer has succeeded if the final state is
\begin{equation}
	\ket{\Psi_{\rm final}} = \ket{O} \left( \alpha \ket{+} + \beta \ket{-} \right),\label{Psifinaldef}
\end{equation}
where $\ket{O}$ is the electron vacuum (all quantum dots empty) and $\ket{\pm}$ represents the two photon states of opposite circular polarization. For later use we define also the states
\begin{eqnarray}
	&&\ket{\Psi_{\pm}} = \alpha \ket{+} \pm \beta \ket{-},\label{Psipmdef}\\
	&&\ket{\Phi_{\pm}} = \left( \ket{A} \pm \ket{B} \right)/\sqrt{2}.\label{Phipmdef}
\end{eqnarray}

The reversible radiative transitions (with rate $g$) are described by the Hamiltonian
\begin{equation}
	H_{g} = g \bigl( \kb{A+}{A'0} + \kb{B-}{B'0} \bigr) + {\rm H.c.}.\label{Hgdef}
\end{equation}
The reversible tunnel transitions (with rate $T$) between dots $A$ or $B$ and dot $C$ have Hamiltonian
\begin{equation}
	H_{T} = T \bigl( \kb{C}{A} + \kb{C}{B} \bigr) \otimes \openone_{\rm photon} + {\rm H.c.},\label{HTdef}
\end{equation}
where $\openone_{\rm photon}$ is the unit operator acting on the photons.

The irreversible escape (with rate $\Gamma$) of the electron from dot $C$ into the reservoir is described by the jump operator
\begin{equation}
	D_{\Gamma} = \sqrt{\Gamma}\, \kb{O}{C} \otimes \openone_{\rm photon}.\label{DGammadef}
\end{equation}
These operators determine the time evolution of the density matrix $\rho(t)$ through the master equation\cite{Naz93,Gur98}
\begin{equation}
	\frac{d\rho}{dt} = -i[H,\rho] + D_{\Gamma}^{\vphantom{\dagger}} \rho D_{\Gamma}^{\dagger} - \tfrac{1}{2} \left( D_{\Gamma}^{\dagger} D_{\Gamma}^{\vphantom{\dagger}} \rho + \rho D_{\Gamma}^{\dagger} D_{\Gamma}^{\vphantom{\dagger}} \right).\label{master}
\end{equation}
(We have set $\hbar$ to $1$.) The initial condition is $\rho(0) = \kb{\Psi_{\rm in}}{\Psi_{\rm in}}$.

Inspection of the master equation shows that the density matrix evolves entirely in the five-dimensional subspace spanned by the states
\begin{eqnarray}
	&&\ket{u_{1}} = \ket{\Psi_{\rm in}},\;\; \ket{u_{2}} = \ket{\Phi_{+}}\ket{\Psi_{+}},\;\; \ket{u_{3}} = \ket{\Phi_{-}} \ket{\Psi_{-}},\nonumber\\
	&&\ket{u_{4}} = \ket{C} \ket{\Psi_{+}},\;\;\ket{u_{5}} = \ket{\Psi_{\rm final}}\label{basisset}
\end{eqnarray}
of even parity under the exchange $\alpha\leftrightarrow\beta$, $A\leftrightarrow B$, $A'\leftrightarrow B'$, $\sigma_{+}\leftrightarrow\sigma_{-}$. The states $(\alpha\ket{A'} - \beta\ket{B'})\ket{0}$, $\ket{\Phi_{+}}\ket{\Psi_{-}}$, $\ket{\Phi_{-}}\ket{\Psi_{+}}$, $\ket{C}\ket{\Psi_{-}}$, and $\ket{O}\ket{\Psi_{-}}$ of odd parity do not appear.

The five-dimensional subspace may be further reduced to a four-dimensional subspace by noting that the master equation (\ref{master}) couples only to $\rho_{55}$ and to $\rho_{ij}$ with $i,j\leq 4$. The matrix elements $\rho_{ij}$ with $i=5,j\neq 5$ or $j=5,i\neq 5$ remain zero. We may therefore seek a solution of the form
\begin{equation}
	\rho(t) = \tilde{\rho}(t) + \left[1-{\rm Tr}\,\tilde{\rho}(t)\right] \kb{\Psi_{\rm final}}{\Psi_{\rm final}},
\end{equation}
where $\tilde{\rho}$ is restricted to the four-dimensional subspace spanned by the states $\ket{u_{i}}$ with $i\leq 4$.

The evolution equation for $\tilde\rho$ reads
\begin{eqnarray}
	&&\frac{d\tilde{\rho}}{dt} = M\tilde{\rho} + \tilde{\rho}M^{\dagger},\\
	&&M = -\frac{1}{\sqrt{2}} \begin{pmatrix} 0&ig&ig&0\\ ig&0&0&2iT\\ ig&0&0&0\\ 0&2iT&0&\Gamma/\sqrt{2} \end{pmatrix}.\label{Mdef}
\end{eqnarray}
The solution
\begin{equation}
	\tilde{\rho}(t) = e^{Mt} \tilde{\rho}(0) e^{M^{\dagger}t}
\end{equation}
has a lengthy expression in terms of the eigenvalues and eigenvectors of the matrix $M$. What is important for deterministic quantum state transfer is that all four eigenvalues $\mu_{i}$ have a negative real part, for any nonzero $g$, $T$, and $\Gamma$. This implies that $\tilde{\rho}(t)\rightarrow 0$ for $t\rightarrow\infty$, so $\rho(t)\rightarrow\kb{\Psi_{\rm final}}{\Psi_{\rm final}}$.

\begin{figure}[htp]
	\includegraphics[angle=-90,width=8cm]{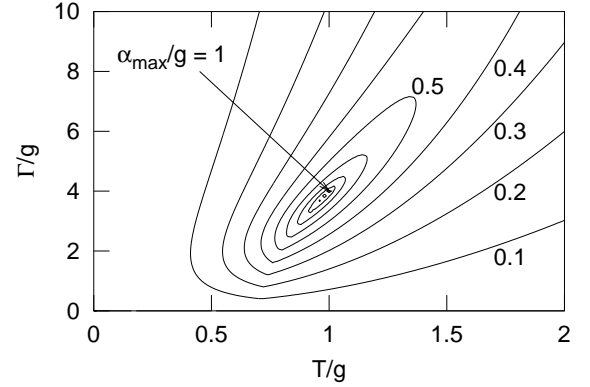}
	\caption{Contour plot of the rate $\alpha$ [defined in Eq.\ (\ref{alpharesult})] at which the fidelity of the quantum state transfer approaches unity, as a function of the tunnel rates $T$ and $\Gamma$. All rates are normalized by the electron-photon coupling constant $g$.
	\label{alpha_plot}}
\end{figure}

The fidelity of the quantum state transfer,
\begin{equation}
	\mathcal{F} = \bok{\Psi_{\rm final}}{\rho}{\Psi_{\rm final}} = 1-{\rm Tr}\,\tilde{\rho},
\end{equation}
approaches unity in the long-time limit with a rate determined by the eigenvalue with real part closest to zero:
\begin{equation}
	\lim_{t\rightarrow\infty} 1 - \mathcal{F}(t) \propto e^{-\alpha t},\;\;\alpha = 2 \min(|{\rm Re}\,\mu_{i}|).\label{alpharesult}
\end{equation}
The asymptotic limits of $\alpha$ are
\begin{equation}
	\alpha = \left\{ \begin{array}{ll} T^2 \Gamma / 2 g^2 & \text{ for } g \gg T,\Gamma\\
	g^2 \Gamma / 8 T^2 & \text{ for } T \gg g,\Gamma\\
	2 T^2 / \Gamma & \text{ for } \Gamma \gg g,T.\end{array}\right.\label{eq:alphalim}
\end{equation}
If one varies $\Gamma$ and $T$ at fixed $g$, the rate $\alpha$ reaches its maximum of $\alpha_{\rm max} = g$ at $T/g = 1, \Gamma/g = 4$. (See Fig.\ \ref{alpha_plot}.) That $\alpha$ vanishes for large $\Gamma,T,g$ can be understood as a manifestation of the quantum Zeno effect:\cite{Mis77} the electron remains trapped in the quantum dots because the decay into the reservoir is inhibited by too frequent measurement. The optimal rate $\alpha_{\rm max} = g$ implies that about 1 optical cycle of emission/absorption of the photon in the cavity is needed to effectively transfer the state. This seems fast enough, in view of the inevitable losses in the cavity.

The fidelity of the quantum state transfer is reduced by decoherence of the charge qubit due to coupling of the charge to acoustic phonons\cite{Wu05,Vor05} or to background charge fluctuations.\cite{Ita03} Following Ref.\ \onlinecite{Gur97} we model this decoherence (with rate $\Gamma_{\phi}$) by means of the jump operators
\begin{eqnarray}
	&&D_{X} = \sqrt{\Gamma_{\phi}} \bigl( \kb{X}{X} + \kb{X'}{X'} \bigr) \otimes \openone_{\rm photon},\nonumber\\
	&&X \in \{A,B\},\;\;D_{C} = \sqrt{\Gamma_{\phi}}\kb{C}{C} \otimes \openone_{\rm photon},\label{DXdef}
\end{eqnarray}
which measure the charge on each of the three dots. The master equation (\ref{master}) now becomes
\begin{eqnarray}
	\frac{d\rho}{dt} &=& -i[H,\rho] + \sum_{X=\Gamma,A,B,C} D_{X}^{\vphantom{\dagger}} \rho D_{X}^{\dagger}\nonumber\\
	&&\mbox{} - \tfrac{1}{2} \sum_{X=\Gamma,A,B,C} \left( D_{X}^{\dagger} D_{X}^{\vphantom{\dagger}} \rho + \rho D_{X}^{\dagger} D_{X}^{\vphantom{\dagger}} \right),\label{masternew}
\end{eqnarray}
where $D_{\Gamma}$ was defined in Eq.\ (\ref{DGammadef}). In what follows we take $\alpha = \beta = \tfrac{1}{2} \sqrt{2}$: since the initial state (\ref{Psiindef}) is then maximally delocalized, it will be most sensitive to decoherence.

\begin{figure}[!htp]
	\includegraphics*[width=8cm]{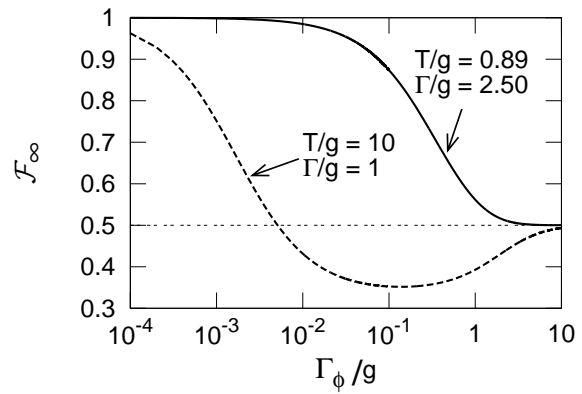}
	\caption{Decay of the long-time fidelity $\mathcal{F}_\infty$ of the quantum state transfer with increasing decoherence rate. We have taken $\alpha = \beta = \tfrac{1}{2}\sqrt{2}$. The solid curve is for parameter values at which the quantum state transfer is least sensitive to decoherence.}
	\label{fig:fidelity}
\end{figure}

Results for $\mathcal{F}_\infty = \lim_{t\rightarrow \infty} \mathcal{F}(t)$ are plotted in Fig.\ \ref{fig:fidelity} for two parameter choices. The asymptotic limits are
\begin{align}
	\label{eq:fidelity}\mathcal{F}_\infty &= \left\{ \begin{array}{ll} 1 - A \Gamma_\phi & \text{ for } \Gamma_\phi \ll \Gamma,T,g\\
	\frac{1}{2}\; +\; B \Gamma_{\phi}^{-4} & \text{ for } \Gamma_\phi \gg \Gamma,T,g \end{array} \right.\\
	A &= \frac{4g^4+16T^4+g^2 \Gamma^2-10 g^2 T^2}{4g^2T^2\Gamma},\label{eq:A}\\
	B &= T^2 \left(3g^2-2T^2\right).
\end{align}
Note that $B$ is independent of $\Gamma$. By comparing the expression (\ref{eq:A}) for the coefficient $A$ with Eq.\ (\ref{eq:alphalim}) for the transfer rate $\alpha$, we see that $\mathcal{F}_\infty = 1 - \Gamma_\phi / 2 \alpha + \mathcal{O}(\Gamma_\phi^2)$ if one of the three rates $\Gamma, g, T$ is much larger than the other two. In this regime the sensitivity to decoherence is determined entirely by how fast the state can be transferred.

As found in Ref.\ \onlinecite{Ema05} in connection with the entanglement production, the effect of decoherence on the charge qubit is minimal if $T \approx \Gamma \approx g$. More precisely, the fidelity $\mathcal{F}_\infty$ is maximized for fixed $g$ and $\Gamma_\phi$ if $T/g = \tfrac{2}{5}\sqrt{5} \approx 0.89,\; \Gamma/g = \tfrac{2}{5}\sqrt{39} \approx 2.50$ if $\Gamma_\phi \ll g$ and if $T/g = \tfrac{1}{2}\sqrt{3} \approx 0.87$ if $\Gamma_\phi \gg g$. As shown in Fig.\ \ref{fig:fidelity} the fidelity depends non-monotonically on $\Gamma_\phi$, approaching the asymptotic limit $\tfrac{1}{2}$ from below for $T/g > \sqrt{3/2} \approx 1.22$. When $\mathcal{F}_\infty < \tfrac{1}{2}$ the fidelity of the quantum state transfer can be improved by exchanging the photon polarizations ($\sigma_+ \leftrightarrow \sigma_-$), so that $\mathcal{F}_\infty \mapsto 1 - \mathcal{F}_\infty$. With this procedure the fidelity may actually \emph{increase} with increasing $\Gamma_\phi$.

In conclusion, we have analyzed a mechanism for the quantum state transfer from charge qubits to photon qubits which is deterministic (no post-selection is required) and which depends only algebraically on the decoherence rate. The mechanism relies on the coupled dynamics of an electron and a photon in a microwave cavity, but the transfer can be sufficiently fast so that only a few optical cycles of emission/absorption are required. Decoherence rates as large as $10\%
$ of the emission rate then do not degrade the fidelity of the quantum state transfer below about $0.9$. These characteristics suggest that the mechanism considered here might have promising applications in quantum information processing.

This research was supported by the Dutch Science Foundation NWO/FOM.


\end{document}